# A gravitational lens model for B1422+231

**R. Kormann**\*, **P. Schneider** and **M. Bartelmann**

Max-Planck-Institut für Astrophysik, Postfach 1523, D–85740 Garching, FRG

November 2, 1993

**Abstract.** We investigate simple gravitational lens models for the recently discovered quadruple QSO B1422+231. The image configuration in this multiple QSO system indicates that the gravitational deflector has a considerable ellipticity; we therefore explore lensing models based on elliptical mass distributions. Employing a simple fitting procedure, we show that (1) our models cannot satisfactorly account for the details of the image configuration, in particular, the flux ratios are difficult to reproduce; (2) our models yield a somewhat better fit than the recently published model by Hogg & Blandford. (3) Our best model is a combination of an elliptical deflector and externally applied shear; the direction of the shear is in accordance with the position of two galaxies detected near the QSO images, and may well be produced by them. If this interpretation is correct, our best model is in fact quite similar to the model of Hogg & Blandford, except that we account for the ellipticity of the main deflector, which reduces the mass of the nearby galaxies to produce the necessary shear. The fact that the models do not improve if the position of the lens center is allowed to vary from its observed position is a strong argument in favour of it being indeed the main lensing galaxy.

**Thesaurus codes:** 02.07.1, 12.07.1

## 1 Introduction

Most gravitational lens systems, in which a distant QSO is multiply imaged by a foreground deflector, can be modelled by 'canonical' gravitational lens models (e.g., Narayan & Grossman 1989; Kochanek 1991). Typically, one can fit the image positions and the approximate flux ratios of the images by a simple 'elliptical' deflector, where it does not matter much whether homoeoidal elliptical mass distributions are used, or spherical models with superimposed shear, or lenses with elliptical isopotentials. Except for the triple QSO 2016+112 (Lawrence et al. 1984), for which simple lens models do not work (R.D. Blandford, private communication, but see Narasimha, Subramanian & Chitre 1987; note also that at least two lensing galaxies, most likely at different redshifts, are involved in imaging this QSO), such modelling by simple lenses does not pose a great challenge.

---

\* Present address: Fraunhofer-Gesellschaft, Kreuzeckbahnstraße 19, D–82467 Garmisch-Partenkirchen, FRG

*Send offprint requests to:* P. Schneider

The recently discovered multiple QSO B1422+231 (Patnaik et al. 1992; hereafter PBWCF) is a gravitational lens system with four images, whose radio positions agree well with the infrared morphology (Lawrence et al. 1992; hereafter LNWMP). The image configuration of the system is close to one that is expected from 'canonical' models, but not quite so, and it may therefore resist simple modelling attempts. In fact, in a recent preprint, Hogg & Blandford (1993; hereafter HB) have shown that the image configuration can be modelled satisfactorily, but the flux ratios of the images obtained from the model disagree substantially from the observed ones.

In this paper, we will investigate whether the disagreement between the HB models and the observations are due to the restricted class of lens models used in HB, or whether it is intrinsically difficult to reproduce the observations from simple, or plausible, lens models. In Sect. 2, we will briefly review some observational results on B1422+231, and we discuss the lens models of HB in Sect. 3. Our lens models are based on the singular 'isothermal' ellipsoidal lens, whose properties have been investigated recently by Kassiola & Kovner (1993; hereafter KK) and Kormann, Schneider & Bartelmann (1993; hereafter KSB). Since the ellipticity of the mass distribution is large (we shall see that we need an axis ratio of $\sim 1 : 2.5$), lenses with elliptical potentials are useless here, since they yield unphysical mass distributions (see Schramm 1990 and KK for further comments on this point). We describe our fitting procedure in Sect. 4, and present our lens models in Sect. 5, where they are compared with the HB models. We shall see that our models do not significantly improve the fit compared to the HB models, but the difficulties we encounter differ from theirs. We discuss our results in Sect. 6. In the appendix, we briefly discuss the lensing properties of the singular isothermal ellipsoid with a superimposed quadrupole perturbation.

## 2 Selected observational results

Radio observations of B1422+231 at 5 GHz (with MERLIN) and 8.4 GHz (with the VLA) were presented in PBWCF. They discovered four components at both frequencies; their positions and flux ratios are reproduced in Table 1. The three brighter components A, B, and C agree in spectral index between these two frequencies, as well as in their degree and direction of the polarization. The spectral index of the fainter D image differs from that of the other three images, but due to its faintness, the uncertainties of the fluxes of the D image are considerable. The infrared observations of LNWMP reveal the same four-image morphology; in particular, they also found the D image, but slightly brighter relative to the others than in the radio maps. The source in B1422 is a QSO with redshift $z_s = 3.62$, which is very bright in the optical, making this an apparently extremely luminous QSO.

As reported in HB, the lensing galaxy (G1) has been observed by E. Ellingson & H.K.C. Yee, and a redshift measurement by F. Hammer et al. yielded the preliminary result $z_d = 0.64$; in addition, two more galaxies (G2, G3) close to the QSO images have been observed by C.R. Lawrence et al.; the positions of the galaxies are listed in Table 2. Patnaik (1993) reported that the QSO shows a large rotation measure.

The image configuration immediately provides qualitative constraints on any lens model. The fact that the three bright images are fairly colinear and much brighter than the D image suggests that the source lies close to a cusp singularity, and that these three images are magnified substantially. However, since the flux ratios of these three images do not follow the scaling relation expected from the lens equation near cusps (Mao 1992; Schneider & Weiss 1992), the source cannot be *very* close to a cusp; hence, the magnification cannot be very large. This implies that the D image must be demagnified, so that it has to lie close to the center of the lensing galaxy; in other words, the source has to be fairly close to the radial caustic of the lens. Together, these two remarks imply that the imaging of B1422+231 is provided by a lens which has a naked cusp singularity, or at least that the cusp is very close to the radial caustic. Therefore, the lens must either have a large ellipticity, or else be perturbed by a large tidal potential, i.e., an external shear.

## 3 The models by Hogg & Blandford

Two lens models for B1422+231 were considered in HB. In the first ('initial') model, which was constructed prior to the measurement of the galaxy positions, two singular isothermal spheres were used. One of the two galaxies is situated between the images and acts as the primary lensing galaxy. The second galaxy lies 7″ away from the images, and its main purpose is to provide the shear. This large separation, and the fact that the deflecting potential must be fairly eccentric, implies that the second galaxy must be very massive; its velocity dispersion must be about 460 km/s. The initial model has six free parameters to characterize the lens, i.e., the two centers of the galaxies and the two values of the velocity dispersion.

The second model ('refined model') uses the observed positions of the galaxies (G1, G2, G3); G1 is modelled as an isothermal sphere, whereas G2 and G3 are treated as point mass lenses. It is assumed that the galaxies G2 and G3 have the same redshift as G1, thus yielding three free lens parameters. The resulting model yields a velocity dispersion of 210 km/s for G1, and masses of $4.1 \times 10^{12} h^{-1} M_\odot$ and $4.7 \times 10^{11} h^{-1} M_\odot$ for G2 and G3, respectively, where $h$ is the Hubble constant in units of 100 km/s/Mpc, and the numbers have been evaluated for an Einstein–de Sitter universe.

The introduction of the refined model improves the fit considerably, compared to the initial model, as measured by $\chi^2/\nu$, where $\nu$ is the number of degrees of freedom. Still, HB's best model has a $\chi^2/\nu$ of 16; most of the difference between the model and the observed image properties is due to the flux ratio between images A and B: the radio observations show $|\mu_A|/|\mu_B| = 0.98 \pm 0.02$, whereas the model predicts this ratio to be 0.76. The infrared observations in LNWMP suggest that the flux ratio between A and B may actually be closer to the value predicted by the model; on the other hand, the flux ratio of 0.98 has been measured at both radio frequencies.

The considerable value of $\chi^2/\nu$ in the refined model of HB, and the fairly large mass for the galaxy G2 it implies, has motivated us to try another family of lens models for B1422+231. Since both models of HB clearly demonstrate the need for a large asphericity of the matter distribution, in accordance with the remarks made at the end of the preceding section, it is natural to model the main lensing galaxy as an elliptical mass distribution. We expect that the required ellipticity will be fairly large, perhaps like that of a spiral galaxy viewed nearly edge-on. Both, the small image splitting and the large rotation measure (Patnaik 1993) indicate that the main lens may indeed be a spiral galaxy.

# 4 Fitting procedure

## 4.1 Coordinates

We introduce coordinates $(x_1, x_2)$ centered on G1, with axes along (decreasing) right ascension and (increasing) declination. The semi-minor axis of the elliptical lens encloses an angle $\phi_g$ with the positive $\theta_1$-axis. Let $\xi_0$ be the length scale in the lens plane; then, the dimensionless coordinates $(x_1, x_2)$ in the lens plane and $(y_1, y_2)$ in the source plane are related to observed angles by the relations

$$\boldsymbol{\theta} = \theta_0 \boldsymbol{x} \quad , \quad \boldsymbol{\beta} = \theta_0 \boldsymbol{y} , \tag{1}$$

where $\theta_0$ is defined by

$$\theta_0 \equiv \frac{\xi_0}{D_\mathrm{d}} . \tag{2}$$

The length scale $\xi_0$ is defined in Eq.(A2), and $D_\mathrm{d} \equiv D(z_\mathrm{d})$ is the angular-diameter distance from the observer to the lens. With the redshifts of the lens (supposed to be G1) and the source, $z_\mathrm{d} = 0.64$ and $z_\mathrm{s} = 3.62$, respectively, from Eqs.(2) and (A2), and from the distance-redshift relation in an Einstein-de Sitter universe, we obtain the relation between the velocity dispersion of the lens and $\theta_0$,

$$\sigma_v = 242 \, \frac{\mathrm{km}}{\mathrm{s}} \, \sqrt{\theta_0} , \tag{3}$$

where from now on, all angles are measured in arcseconds. The positions of the images of B1422+231, and their fluxes relative to the brightest image (B) are summarized in Tab.1.

**Table 1.** Image coordinates and flux ratios of B1422+231. The coordinate origin is taken to be the position of G1, the coordinates are given in arcseconds. Fluxes are related to the brightest image (B)

| Image | $\theta_1^\mathrm{o}$ | $\theta_2^\mathrm{o}$ | $\delta^\mathrm{o}$ |
|-------|------|------|------|
| A | $0.31 \pm 0.02$ | $0.91 \pm 0.02$ | $0.98 \pm 0.02$ |
| B | $0.70 \pm 0.02$ | $0.59 \pm 0.02$ | $1.00 \pm 0.02$ |
| C | $1.03 \pm 0.02$ | $-0.16 \pm 0.02$ | $0.52 \pm 0.02$ |
| D | $-0.24 \pm 0.03$ | $-0.22 \pm 0.03$ | $0.02 \pm 0.005$ |

**Table 2.** Coordinates of the galaxies G2,G3, relative to G1, in arcseconds

| galaxy | $\theta_1$ | $\theta_2$ |
|--------|------|------|
| G1 | 0 | 0 |
| G2 | $-8.3$ | $-4.6$ |
| G3 | $-2.9$ | $-6.7$ |

## 4.2 Parameters

We will investigate three different gravitational lens models for the images in the system B1422+231. All three models are based on the singular isothermal ellipsoid, as investigated in KK and KSB, using the notation of the latter. The singular isothermal ellipsoid is specified by three parameters, the axis ratio $f$ ($\leq 1$), the position angle $\phi_g$ of its semi-minor axis, and its 'strength', which can be expressed in terms of the length scale $\xi_0$ incorporated in $\theta_0$ of Eq.(2). In the model 1, we choose the lensing galaxy to lie at the position of galaxy G1. Allowing for the fact that the lens position may be determined inaccurately, the position of the center of the lensing galaxy is a free parameter in model 2, and we denote the position of the lens center relative to the observed position of G1 by $\boldsymbol{R}$. In model 3 we put the lens center at the position of G1, but allow for some external shear; hence, we add to the deflection angle caused by the galaxy a tidal term

$$\boldsymbol{\alpha}_s(\boldsymbol{\theta}) = \gamma \begin{pmatrix} -\cos 2\phi_s & -\sin 2\phi_s \\ -\sin 2\phi_s & \cos 2\phi_s \end{pmatrix} \boldsymbol{\theta} \;, \tag{4}$$

where $\gamma$ describes the strength of the shear, and $\phi_s$ its direction. The physical meaning of the direction is such that a point mass at position angle $\phi_s$ would produce a tidal term of the form (4), with $\gamma > 0$. Model 3 accounts for the possibility that the line-of-sight to the QSO may be perturbed by masses at the same or a different redshift as the main lens, e.g., by the galaxies G2 and G3.

The total number of observables is 12, namely 4×2 for the coordinates of the images, and 4 for the fluxes of the images. The number of parameters at hand is 6 for model 1 (i.e., the three lens parameters of the isothermal ellipsoid, one for the intrinsic 'luminosity' of the source, and the two source coordinates). Models 2 and 3 have two additional parameters, the lens displacement $\boldsymbol{R}$ and the shear parameters $\gamma$, $\phi_s$, respectively. Thus, we have 6 degrees of freedom for model 1, and 4 for models 2 and 3.

In addition to the three models described above, we have also considered a lens model similar to the refined model of HB, taking the observed positions of the three galaxies G1, G2, and G3, and modelling the first as a singular isothermal sphere, and the latter two as point mass lenses. The lens parameters are named as in HB, i.e., $b_1$ denotes the radius of the Einstein circle of G1, and $b_2$ and $b_3$ are the Einstein radii of the point mass lenses corresponding to G2 and G3, respectively, all given in arcseconds. In this model, which we henceforth call HB model, the number of degrees of freedom is 6. The model parameters are summarized in Table 3. For each of the four models just mentioned, we have tried to match the observations, as described in the following subsection, minimizing the least-square merit function (6). By definition, it contains the measurement errors of the image positions and fluxes. We consider two sets of values for these errors: in the first set, the errors are those given in Table 1 and agree with those used in HB. In the second set, the error of the flux measurement of image D relative to image B is increased to 0.02, with all other errors unchanged. We denote the resulting models in the former case with suffix 'a', and the latter ones with suffix 'b'. Hence, model 2a denotes model 2 with the measurement errors as given in Table 1.

**Table 3.** Model parameters

| parameter | symbol | used in models |
|---|---|---|
| lens 'strength' | $\theta_0$ | 1,2,3 |
| axis ratio | $f$ | 1,2,3 |
| position angle of ellipticity | $\phi_g$ | 1,2,3 |
| lens displacement | $(R_1, R_2)$ | 2 |
| source position | $(\beta_1, \beta_2)$ | all |
| source 'luminosity' | $l$ | all |
| external shear | $\gamma$ | 3 |
| position angle of external shear | $\phi_s$ | 3 |
| HB lens 'strengths' | $(b_1, b_2, b_3)$ | HB |

### 4.3 Fitting procedure

The model fit proceeds in three steps:

1. For a given set of model parameters, we determine four positions $\boldsymbol{\beta}_i$, $1 \leq i \leq 4$ in the source plane by inserting the observed image positions into the corresponding lens equation, i.e., $\boldsymbol{\beta}_i = \boldsymbol{\theta}_i^o - \boldsymbol{\alpha}(\boldsymbol{\theta}_i^o)$, using the lens equation as given in Eq.(A6) of the Appendix. These four positions are weighted with the observed relative image fluxes $\delta_i^o$ according to the method described by Kochanek (1991),

$$\boldsymbol{\beta} = \frac{\sum (\delta_i^o)^2 \boldsymbol{\beta}_i^o}{\sum (\delta_i^o)^2} ,  \qquad (5)$$

   where the sum extends over the four images. This value of $\boldsymbol{\beta}$ is henceforth used as the source position, i.e., in the process of finding the best model, the source position is not varied any more. Hence, our model procedure will not find the best possible model parameters, but the procedure used here guarantees a fast performance of the fitting. Since our resulting models are not 'acceptable' from the $\chi^2/\nu$-point of view, a more refined modelling seems not warranted at this stage.

2. Image positions $\boldsymbol{\theta}_i$ and image magnifications (and magnification ratios $\delta_i$ relative to image B) from this source position are then computed using the lens equation (A7).

3. Given the reconstructed image positions, the reconstructed fluxes, and the model parameters, we compute the corresponding least-square merit function $\chi^2$ defined by

$$\chi^2 = \sum \left[ \frac{(\boldsymbol{\theta}_i^o - \boldsymbol{\theta}_i)^2}{\sigma_i^2} + \frac{(l\delta_i - \delta_i^o)^2}{(\sigma_i^f)^2} \right] . \qquad (6)$$

   Where not otherwise stated, we use the values given in Tab.1 for the measurement errors in position $\sigma$ and in flux $\sigma^f$; i.e., $\sigma = 0.02 = \sigma^f$ for images A,B, and C, and $\sigma = 0.03$, $\sigma^f = 0.005$ for image D. This $\chi^2$ is then minimized varying the model parameters employed. Note that the $\chi^2$ in Eq.(6) is not minimized varying all model parameters including the source position, but that the source position is determined from Eq.(5) and later kept fixed, as already mentioned above. This, however, does not affect the number of degrees of freedom, since Eq.(5) imposes two restrictive relations between observed and modelled quantities.

# 5 Results

## 5.1 Models 1a through 3a, model HBa

For the first set of models, we apply the fitting procedure described above, using the observational errors as listed in Table 1. The resulting lens models are listed in Table 4 and displayed in Fig. 1.

The first point to note is that none of the resulting models has a satisfactorly small value of $\chi^2/\nu$, i.e., none of the models can reproduce the observed image properties. Although models 2a and 3a have a slightly lower value of $\chi^2/\nu$ than the model HBa, this is not a sufficient improvement to consider these models clearly superiour to the HB model. We would like to note that our fitting procedure has yielded a slightly larger value for $\chi^2/\nu$ for the HBa model than quoted in HB, due to the fact that we have not optimized the source position. The largest deficit of the HBa model is the flux ratio of image A, relative to image B, as also concluded by HB.

The worst of our models is model 1a, where the flux of image C, its position, and the flux of image D is badly reproduced. The problem with that model is quite different from that of the HB model, in that the flux ratio between A and B is very well matched with the observations. The fact that image D is too bright in the model stems from the fairly large separation of the image from the center of the lensing galaxy G1: in order for image D to be sufficiently demagnified it has to lie closer to the lens position. This fact can be easily seen by considering model 2a, where the lens position is a free parameter. The best fit model is one where the lens position is shifted towards the image D, yielding a sufficient demagnification to match the observed flux ratios. Also for this model, the flux of image C is too large. In both of these models, the axis ratio is large, about $1:2.5$ (corresponding to $f \sim 0.4$), which is in accord with the remarks made at the end of Sect. 2. If we allow for an external shear, as in model 3a, the axis ratio of the galaxy can be reduced (i.e., $f$ be increased), since the shear is oriented such that the asymmetry of the lens mapping is increased: external shear and galaxy ellipticity act in nearly the same direction. Model 3a yields a fairly good reproduction of the positions and fluxes of the three bright images, but does not yield a good match of the properties of image D.

## 5.2 Models 1b through 3b, model HBb

The fact that the largest problem of our models was to fit the flux of image D, we have decided to repeat the model fitting, this time with an increased error of the flux of image D, i.e., by choosing $\sigma_D^f = 0.02$ instead of $0.005$. This choice is motivated by the fact that image D is the faintest, so that we consider the measurement inaccuracies to be potentially largest for this image. In addition, the best fit models of the infrared data of LNWMP yield a flux ratio of image D relative to B of 0.06. The results of this second set of models are listed in Table 5 and displayed in Fig. 2.

Comparing these new models with the former ones shows that all models have a drastically improved value of $\chi^2/\nu$, except the HB model; the reason why the HB model does not improve markedly is seen from Fig. 1 which shows that the flux of image D does not contribute significantly to $\chi^2$ in this model. Also, model 2 does not improve dramatically, since the fact that the galaxy position is a free parameter in this model gets rid of the problem of an overly luminous image D, by shifting the galaxy position close to this image. However, models 1b and 3b yield (formally, at least) a much better fit to the observed image properties than the corresponding models 1a and 3a. Also, in going from the 'a' models to the 'b' model, the ellipticities of the deflector are reduced in models 1 and 2, and the amplitude of the external shear in model 3 is reduced, whereas the orientation of the galaxy and the direction of the shear stay nearly constant.

**Table 4.** First set of models: best-fit model parameters, magnification factors, total magnification $\mu_\mathrm{t}$, and $\chi^2/\nu$

| Model #<br>parameter | 1a | 2a | 3a | HBa |
|---|---|---|---|---|
| $\theta_0$ | 0.74 | 0.86 | 0.72 | – |
| $f$ | 0.41 | 0.35 | 0.60 | – |
| $\phi_\mathrm{g}$ | $-0.94$ | $-0.98$ | $-1.00$ | – |
| $\delta_\mathrm{A}$ | 0.99 | 1.00 | 0.89 | 0.80 |
| $\delta_\mathrm{C}$ | 0.68 | 0.62 | 0.55 | 0.40 |
| $\delta_\mathrm{D}$ | 0.06 | 0.03 | 0.04 | 0.02 |
| $l$ | 0.95 | 0.97 | 0.96 | 1.11 |
| $R_1$ | – | $-0.23$ | – | – |
| $R_2$ | – | $-0.12$ | – | – |
| $\gamma$ | – | – | 0.16 | – |
| $\phi_\mathrm{s}$ | – | – | $-0.93$ | – |
| $b_1$ | – | – | – | 0.74 |
| $b_2$ | – | – | – | 5.04 |
| $b_3$ | – | – | – | 2.01 |
| $\mu_\mathrm{A}$ | 4.67 | 4.83 | 5.32 | 10.18 |
| $\mu_\mathrm{B}$ | $-4.72$ | $-4.86$ | $-6.01$ | $-12.83$ |
| $\mu_\mathrm{C}$ | 3.22 | 3.03 | 3.28 | 5.06 |
| $\mu_\mathrm{D}$ | $-0.30$ | $-0.12$ | $-0.22$ | $-0.33$ |
| $\mu_\mathrm{t}$ | 12.91 | 12.84 | 14.83 | 28.20 |
| $\chi^2/\nu$ | 26.4 | 13.2 | 13.8 | 19.1 |

# 6 Discussion

In this paper we have investigated the gravitational lens system B1422+231 by trying to fit the image configuration with a set of simple lens models. The models chosen here are based on the singular isothermal ellipsoid, which has simple lensing properties, investigated in KK and KSB. We have used the observed position of the lensing galaxy as a constraint in our models, although we also checked whether the model fits improve

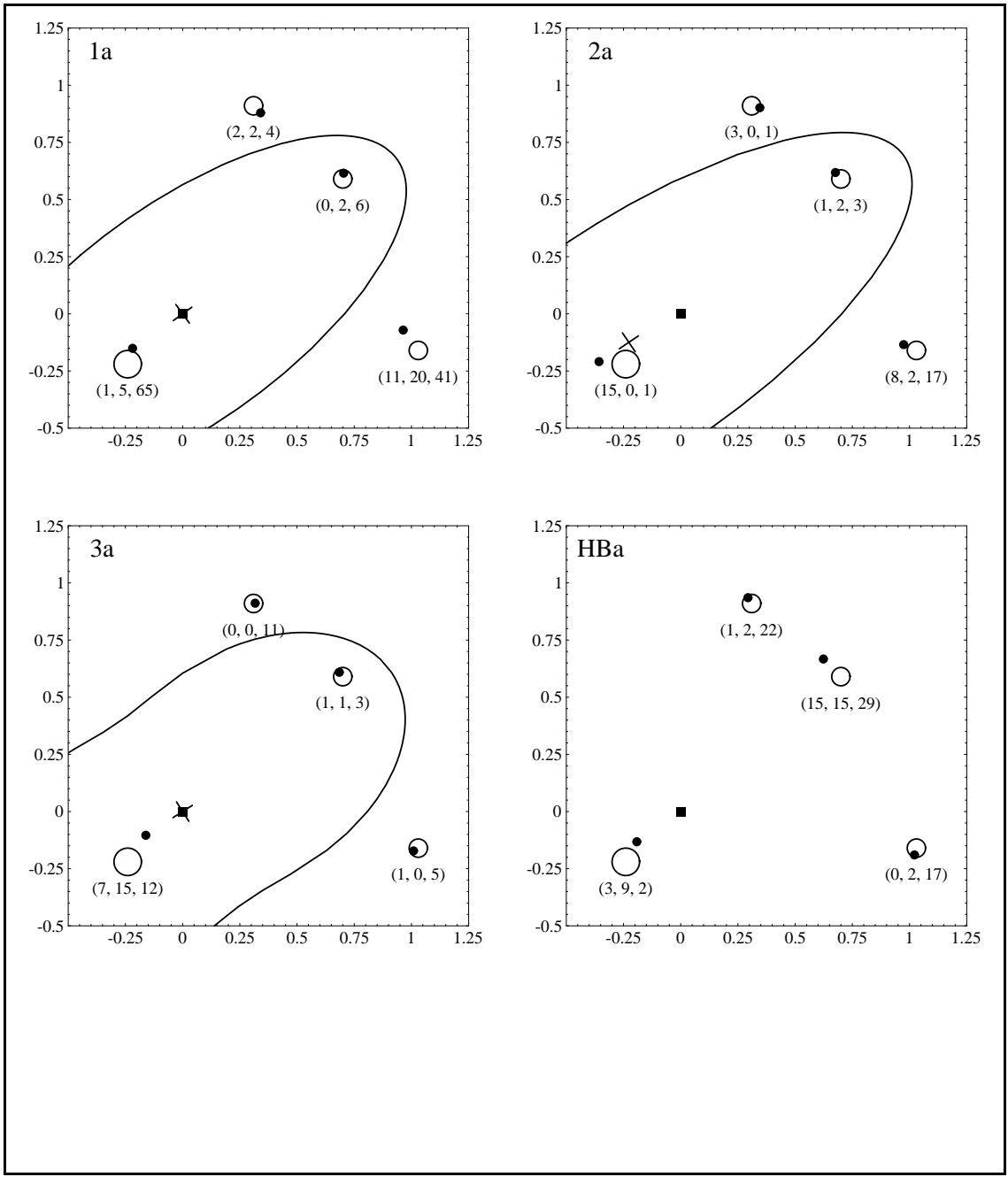

**Fig. 1.** Maps of the observed and the modelled image positions of B1422+231, obtained from models 1a, 2a, 3a, and from model HBa. The open circles show $2\sigma$ error circles around the observed positions, the filled circles the modelled positions, the filled square is the position of G1, the cross the position of the lens. Also indicated is the critical curve. The triples attached to each image position contain the contributions of that image to $\chi^2$ in the order (position 1, position 2, flux); thus, a high value in one of these places means that the corresponding image property is badly modelled

10                                                                A gravitational lens model for B1422+231**Table 5.** Second set of models: best-fit model parameters, magnification factors, total magnification $\mu_\mathrm{t}$, and $\chi^2/\nu$

| Model #  parameter | 1b    | 2b    | 3b    | HBb    |
|---|---|---|---|---|
| $\theta_0$   | 0.78  | 0.83  | 0.76  | –      |
| $f$          | 0.49  | 0.41  | 0.57  | –      |
| $\phi_\mathrm{g}$ | −0.96 | −0.97 | −1.02 | –      |
| $\delta_\mathrm{A}$ | 0.95  | 0.98  | 0.90  | 0.80   |
| $\delta_\mathrm{C}$ | 0.61  | 0.61  | 0.57  | 0.40   |
| $\delta_\mathrm{D}$ | 0.09  | 0.05  | 0.06  | 0.03   |
| $l$          | 0.98  | 0.98  | 0.98  | 1.11   |
| $R_1$        | –     | 0.12  | –     | –      |
| $R_2$        | –     | −0.06 | –     | –      |
| $\gamma$     | –     | –     | 0.10  | –      |
| $\phi_\mathrm{s}$ | –  | –     | −0.83 | –      |
| $b_1$        | –     | –     | –     | 0.75   |
| $b_2$        | –     | –     | –     | 4.96   |
| $b_3$        | –     | –     | –     | 1.99   |
| $\mu_\mathrm{A}$ | 5.67  | 5.20  | 5.50  | 10.27  |
| $\mu_\mathrm{B}$ | −5.94 | −5.30 | −6.08 | −12.83 |
| $\mu_\mathrm{C}$ | 3.65  | 3.23  | 3.43  | 5.06   |
| $\mu_\mathrm{D}$ | −0.52 | −0.26 | −0.37 | −0.34  |
| $\mu_\mathrm{t}$ | 15.78 | 13.99 | 15.38 | 28.50  |
| $\chi^2/\nu$ | 9.7   | 12.2  | 7.3   | 18.7   |

if the lens center is allowed to shift. The results of our models are summarized in Tables 4 & 5 and Figs. 1 & 2.

Of particular interest to us was the question whether the drawbacks of the models obtained by HB are due to their particular choice of the lensing models, or whether B1422+231 is a system for which modelling is more difficult than for most other multiple-image QSO system. The major problems of the HB model are the 'bad' match of the flux ratio of image A to B, and the fact that they require the galaxy G2 to have a huge mass of $\sim 4 \times 10^{12} h^{-1} M_\odot$. This large mass was needed to provide sufficient shear to the main lensing galaxy in order to produce a highly asymmetrical gravitational lens mapping, which is needed to produce the observed image configuration. Expressing their lens mass in terms of a shear parameter, this yields an equivalent of $\gamma \sim 0.25$.

The use of an elliptical deflector removes the necessity for large external shear. The largest problem in our 'a' models is the flux ratio of image D relative to image B, unless the lens position is allowed to be moved towards image D (as in model 2). Our 'b' models provide an improvement over the 'a' models and over the HB model.

Whereas some of our models, in particular 1b and 3b, provide a substantially improved fit over that of the HB model, it may be premature to take this as evidence for the applicability of these lensing models. In fact it seems that the lens system B1422+231

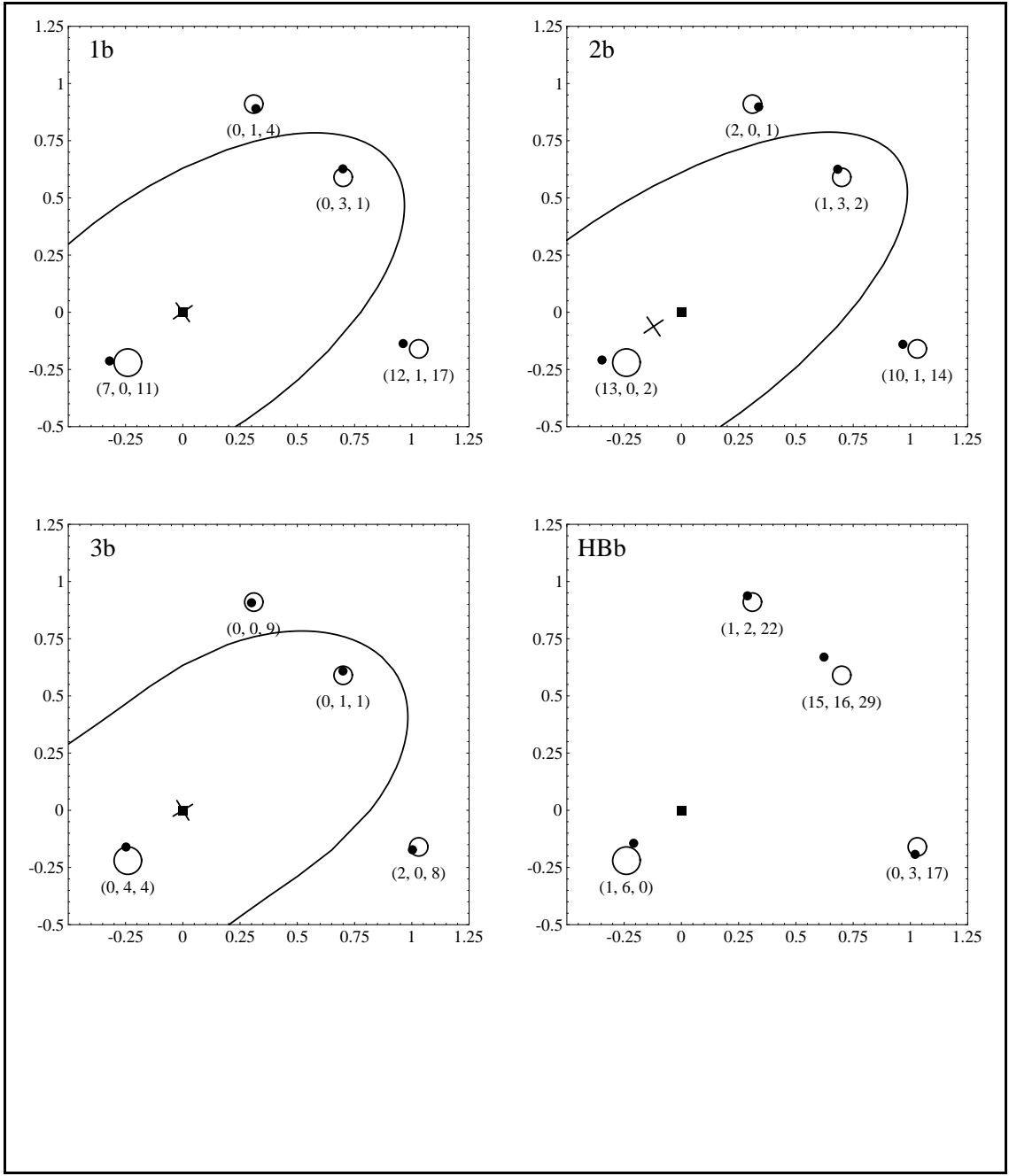

**Fig. 2.** Maps of the observed and the modelled image positions of B1422+231, obtained from models 1b, 2b, 3b, and from model HBb. The open circles show $2\sigma$ error circles around the observed positions, the filled circles the modelled positions, the filled square is the position of G1, the cross the position of the lens. Also indicated is the critical curve. The triples attached to each image position contain the contributions of that image to $\chi^2$ in the order (position 1, position 2, flux); thus, a high value in one of these places means that the corresponding image property is badly modelled

is not formed by a simple lens model, but that the lens mapping may be irregular on scales comparable to the image separation. Reassuring, however, is the fact that the best model, 3b, is in fact very similar to the HB model, in terms of the assumptions. As it turns out, *the shear in model 3b has the direction as to come from a mass concentration in the direction of the two galaxies G2 and G3.* Thus, these two galaxies may indeed contribute substantially to the lens mapping. The amount of shear required in our model, $\gamma = 0.1$, is smaller than that needed in the HB model, due to the fact that part of the lensing asymmetry is due to the elliptic lens G1. In other words, if we assume that the two galaxies G2 and G3 produce the shear implied in model 3b, we reduce their mass by a factor of $\sim 2.5$ relative to the values quoted by HB.

The fact that (at least for the 'b' models) the fit does not improve by allowing the lens center to be shifted from the observed position of galaxy G1 is a strong indication that G1 is indeed the main lensing galaxy. We also want to point out that the mass of the main lensing galaxy in our models is comparable to that obtained from the HB models.

As can be seen from Tables 4 & 5, the total magnifications in our models are typically smaller by a factor of two compared to HB. This can be traced back to the fact that the ellipticity of our main deflector leads to more elongated critical curves, so that the images are allowed to lie further away from them than in the HB models. This, in turn, means that the source can be further away from the cusp, so that the universal scaling relation of magnification close to cusps can be more easily violated, as is observed. We therefore find it easier to fit the flux ratios of the three bright images. On the other hand, the smaller total magnification yields a smaller magnification bias, but the statistical relevance of this fact can hardly be evaluated.

In accordance with the discussion in HB, our models naturally predict large ratios of the eigenvalues of the magnification tensors at the bright image positions. Typically, this ratio is 6 for images A and B for our best-fit models. Therefore, we expect large image distortions to be seen by future VLBI observations; in particular, we expect large apparent velocities if the compact structure of the source is varying in time. For further discussion of this issue, see HB.

# Appendix

In this appendix, we investigate some aspects of the singular isothermal ellipsoid (SIE) gravitational lens model with superimposed quadrupole perturbation, using the same notation as in KSB. The deflection potential of the SIE with axis ratio $f$ ($f \leq 1$) and minor axis along the $x_1$-axis is

$$\psi_g(\boldsymbol{x}) = x\, \tilde\psi(\phi) = x \frac{\sqrt{f}}{f'} \\ \times \left[ \sin\phi \arcsin(f' \sin\phi) + \cos\phi \operatorname{arcsinh}\left(\frac{f'}{f}\cos\phi\right) \right], \quad (A1)$$

where we use polar coordinates $\boldsymbol{x} = (x\cos\phi, x\sin\phi)$, $f' = \sqrt{1-f^2}$, and $\boldsymbol{x}$ is the impact vector of a light ray in units of the 'natural' length scale

$$\xi_0 = 4\pi \frac{\sigma_v^2}{c^2} \frac{D_{\rm d} D_{\rm ds}}{D_{\rm s}} \quad . \tag{A2}$$

Here, $\sigma_v$ is the one-dimensional velocity dispersion of the lens, and the $D$'s have their usual meaning. The deflection potential $\psi_{\rm g}(\boldsymbol{x})$ satisfies the Poisson equation

$$\nabla^2 \psi_{\rm g}(\boldsymbol{x}) = 2\kappa(\boldsymbol{x}) = \frac{\sqrt{f}}{x\,\Delta(\phi)} \quad , \tag{A3}$$

where $\nabla^2$ is the Laplacian, and $\Delta(\phi) = \sqrt{\cos^2\phi + f^2 \sin^2\phi}$. Writing the Laplacian in polar coordinates, one finds from (A1) and (A3) that

$$\tilde{\psi}''(\phi) + \tilde{\psi}(\phi) = 2x\kappa = \frac{\sqrt{f}}{\Delta(\phi)} \quad . \tag{A4}$$

If we perturb the lens by a quadrupole term (see, e.g., Schneider, Ehlers & Falco 1992, Sect. 8.2), the deflection potential becomes

$$\begin{aligned}\psi(\boldsymbol{x}) &= \psi_{\rm g}(\boldsymbol{x}) + \psi_{\rm p}(\boldsymbol{x}) \\ &= x\,\tilde{\psi}(\phi - \phi_0) + \frac{x^2}{2}\left(\kappa_0 + \gamma_0 \cos 2\phi\right) \;,\end{aligned} \tag{A5}$$

where we now have chosen coordinates in which the perturbation matrix is diagonal and in which the minor axis of the SIE encloses an angle $\phi_0$ with the $x_1$-axis. $\kappa_0$ is the (dimensionless) surface mass density associated with the perturbation, and $\gamma_0$ is the shear. The lens equation $\boldsymbol{y} = \boldsymbol{x} - \boldsymbol{\alpha}(\boldsymbol{x})$ with $\boldsymbol{\alpha}(\boldsymbol{x}) = \nabla\psi(\boldsymbol{x})$ becomes

$$\begin{aligned} y_1 &= (1 - \kappa_0 - \gamma_0)x\cos\phi - a_1(\phi) \;, \\ y_2 &= (1 - \kappa_0 + \gamma_0)x\sin\phi - a_2(\phi) \;, \end{aligned} \tag{A6a}$$

where

$$\begin{aligned} a_1(\phi) &:= (1 - \kappa_0 - \gamma_0)x\cos\phi \\ &\quad - \tilde{\psi}(\phi - \phi_0)\cos\phi + \tilde{\psi}'(\phi - \phi_0)\sin\phi \;, \\ a_2(\phi) &:= (1 - \kappa_0 + \gamma_0)x\sin\phi \\ &\quad - \tilde{\psi}(\phi - \phi_0)\sin\phi - \tilde{\psi}'(\phi - \phi_0)\cos\phi \;. \end{aligned} \tag{A6b}$$

The lens equation can be transformed to a one-dimensional equation, which makes its solution particularly simple. If we multiply the first of Eqs. (A6) with $(1 - \kappa_0 + \gamma_0)\cos\phi$, the second with $(1 - \kappa_0 - \gamma_0)\sin\phi$, and add the two resulting equations, we obtain

$$x = \frac{y_1 + a_1(\phi)}{(1 - \kappa_0 - \gamma_0)}\cos\phi + \frac{y_2 + a_2(\phi)}{(1 - \kappa_0 + \gamma_0)}\sin\phi \quad . \tag{A7}$$

Similarly, if we multiply the first of Eqs. (A6) with $(1 - \kappa_0 + \gamma_0)\sin\phi$, the second with $(1 - \kappa_0 - \gamma_0)\cos\phi$, and take the difference, we get

$$\begin{aligned} 0 &= (1 - \kappa_0 + \gamma_0)[y_1 + a_1(\phi)]\sin\phi \\ &\quad - (1 - \kappa_0 - \gamma_0)[y_2 + a_2(\phi)]\cos\phi \quad . \end{aligned} \tag{A8}$$

Hence, one can (numerically) invert the lens equation by solving (A8), which depends solely on $\phi$, and then find the corresponding values of $x$ from (A7); by inserting these values into the lens equation, one can check whether they are real or spurious solutions.

The Jacobian of the mapping (A6) can be easily calculated by differentiation and using (A4):

$$A = \frac{\partial \boldsymbol{y}}{\partial \boldsymbol{x}} = \begin{pmatrix} 1 - \kappa_0 - \gamma_0 - 2\kappa \sin^2 \phi & \kappa \sin 2\phi \\ \kappa \sin 2\phi & 1 - \kappa_0 + \gamma_0 - 2\kappa \cos^2 \phi \end{pmatrix} . \tag{A9}$$

Its determinant is

$$\det A = \mu^{-1} = (1 - \kappa_0)^2 - \gamma_0^2 - 2\kappa \left(1 - \kappa_0 - \gamma_0 \cos 2\phi\right) . \tag{A10}$$

The critical curve, defined as $\det A = 0$, becomes

$$x = \frac{\sqrt{f} \left(1 - \kappa_0 - \gamma_0 \cos 2\phi\right)}{\left[(1 - \kappa_0)^2 - \gamma_0^2\right] \Delta(\phi - \phi_0)} . \tag{A11}$$

The corresponding caustic can be easily obtained by inserting (A11) into the lens equation (A6). Since the deflection angle at the center of the lens, i.e., at $\boldsymbol{x} = 0$, is unchanged by the perturbation, the cut of this perturbed lens is the same as that of the unperturbed SIE. Note that for $\kappa_0 = 0 = \gamma_0$, these equations tend toward those of the unperturbed SIE as treated in KSB and KK.